# Integrated Computation and Communication with Fiber-optic Transmissions

Jiahao Zhang[1,2], Lu Zhang[1,2*], Xiaodan Pang[1,3,4], Oskars Ozolins[3,4,5], Qun Zhang[2*], Xianbin Yu[1*]

[1] *College of Information Science and Electrical Engineering, Zhejiang University, Hangzhou, 310027, China*
[2] *Shandong Zhike Intelligence Computing Inc., Jinan 250101, China*
[3] *Institute of Photonics, Electronics and Telecommunications, Riga Technical University, 1048 Riga, Latvia*
[4] *RISE Research Institutes of Sweden, 16440 Kista, Sweden*
[5] *Department of Applied Physics, KTH Royal Institute of Technology, 106 91 Stockholm, Sweden*
*zhanglu1993@zju.edu.cn; *qun.zhang@zkictech.com; *xyu@zju.edu.cn

***Abstract:*** *Fiber-optic transmission systems are leveraged not only as high-speed communication channels but also as nonlinear kernel functions for machine learning computations, enabling the seamless integration of computational intelligence and communication.*

***Keywords:*** *Optical neural networks and reservoir computing, Photonics in neuromorphic computing and machine learning devices, Optical nonlinear activation devices.*

## INTRODUCTION

Over the past few decades, the field of communication has undergone remarkable transformations, driven by advancements in network architecture and transmission technologies. Simultaneously, the emergence of machine learning (ML) has revolutionized various industries, paving the way for intelligent communication networks [1]. As outlined in the IMT-2030 framework [2], the future of communication systems lies in the seamless integration of ML with communication technologies, a development that is expected to redefine the capabilities of these networks. This integration is particularly critical for applications like semantic communication, which demand a unified approach to bridging the physical and cyber domains. This convergence has given rise to a new research frontier: integrated sensing, communication, and computation [3].

Despite the central role of fiber-optic technology in modern data transmission, its potential for integrating computational tasks remains largely untapped. The concept of Integrated Computation and Communication (ICAC) within fiber-optic systems presents significant advantages, including improved energy efficiency and optimized bandwidth utilization. By leveraging the inherent properties of optical fibers, ICAC could unlock new possibilities for intelligent communication networks.

In this paper, we explore the feasibility of achieving ICAC using fiber-optic systems. Optical fibers are not merely low-loss communication channels; they also exhibit both linear and nonlinear properties due to the interplay between chromatic dispersion (CD) and the Kerr effect. These properties can be utilized as a nonlinear kernel function for ML tasks. To illustrate this, we conduct a numerical analysis of a 7×80 km fiber transmission link, where speech signals are transmitted through the channel while simultaneously performing a voice recognition task. This task serves as a benchmark to evaluate the nonlinear mapping capabilities of the fiber-based kernel function [4]. Remarkably, the fiber-optic system achieves a recognition accuracy of 90.6% after transmission, a performance level comparable to state-of-the-art photonic computing systems. This study highlights the potential of fiber-optic systems to serve dual purposes—communication and computation—while maintaining high efficiency and accuracy. By bridging the gap between these domains, our work contributes to the broader vision of intelligent, integrated communication networks for the future.

**PRINCIPLES OF THE ICAC SYSTEM**

*Fiber Kernel Function for Computation*

ML algorithms require nonlinear operations, which play a crucial role in transforming the original data into a higher-dimensional space to simplify complex patterns and relationships [4]. Achieving nonlinearity in photonic computing has historically been a challenging problem, primarily due to the optical power attenuation introduced by nonlinear layers [5]. However, the Kerr effect in optical fibers provides a distributed mechanism for implementing nonlinearity, thereby mitigating the attenuation typically associated with incorporating nonlinear operations.

To better understand how the Kerr effect can serve as a source of nonlinearity in ML algorithms, we begin by presenting the nonlinear Schrödinger equation (NLSE) [6], which governs light wave transmission in optical fibers:

$$\frac{\partial A(t,z)}{\partial z} = -\frac{\alpha}{2} A(t,z) - j\frac{\beta_2}{2}\frac{\partial^2 A(t,z)}{\partial t^2} + j\gamma |A(t,z)|^2 A(t,z) \qquad (1)$$

where $A$ represents the envelope of the optical field, $z$ is the propagation distance, $t$ is the time coordinate, $j$ is the imaginary unit, $\alpha$ is the attenuation coefficient, $\beta_2$ is the CD coefficient, and $\gamma$ is the nonlinear coefficient.

The NLSE defines a specific mapping between $A$, $z$, and $t$. Let the original information be encoded in the time domain, represented by $A(t,0)$. At the propagation distance $z$, the signal $A(t,z)$ undergoes both linear and nonlinear effects governed by the NLSE and the distance $z$. This process can be expressed as $A(t,z)=f_z(A(t,0))$, where $f_z$ represents the fiber's nonlinear kernel function.

Since an analytical solution to the NLSE does not exist, an explicit expression for $f_z$ is also unavailable. To better evaluate the effects of the fiber kernel function, we can adopt the idea from split-step Fourier method (SSFM) [7]. In this method, the optical fiber is treated as a series of small segments, with the linear and nonlinear effects applied independently to each segment, as illustrated in Fig. 1(a). The equation describing light evolution after applying the SSFM can be written as:

$$A(t, z + \Delta z) = \exp(N\Delta z)\text{IFT}(\exp(D\Delta z)\text{FT}(A(t,z))) , \qquad (2)$$

where $\Delta z$ is the step size of the SSFM, $D=(j\beta_2\omega^2-\alpha)/2$ is the linear operator, capturing attenuation and CD. $N=j\gamma|A|^2$ is the nonlinear operator, representing the Kerr effect, and $\omega$ is the angular frequency. FT and IFT represent the Fourier transform and its inverse, respectively.

In the conventional perspective of optical fiber communication, CD induces inter-symbol interference (ISI), and the Kerr effect distorts the signal amplitude, introducing nonlinearities. Both of these phenomena degrade the signal, leading to a deterioration in communication performance.

*When the fiber is viewed as a special type of nonlinear kernel function, the ISI caused by CD can be viewed as the internal connection between data, and the distortion caused by Kerr effect represents a nonlinear mapping, providing enriched dynamics that are beneficial for ML algorithms* [8]. *These two effects eventually construct a neural-network-liked structure, which can be viewed as a fiber kernel function.*

Following this complex kernel mapping, the received signal at the receiver side can be processed using simple linear methods, such as ridge regression, to construct a powerful computational system [9]. In this way, the fiber-optic system enables the execution of complex computational tasks. When ridge regression is used for further ML processing, the data after kernel mapping is collected to construct a state matrix **C**. This matrix is then multiplied by the readout matrix $\mathbf{W}_{out}$ to obtain the network output: $\mathbf{Y} = \mathbf{W}_{out} \cdot \mathbf{C}$, and the training process is performed by $\mathbf{W}_{out} = \mathbf{Y} \cdot \mathbf{C}^T \cdot (\mathbf{C} \cdot \mathbf{C}^T + \lambda \mathbf{I})^{-1}$, where $\lambda$ is the regularization factor. Ridge regression, being a simple linear algorithm, is easy to implement on programmable platforms, making it well-suited for integration with the nonlinear fiber kernel. The full operation flow of the fiber kernel function for ML task, taking voice recognition as an example, is shown in Fig. 1.

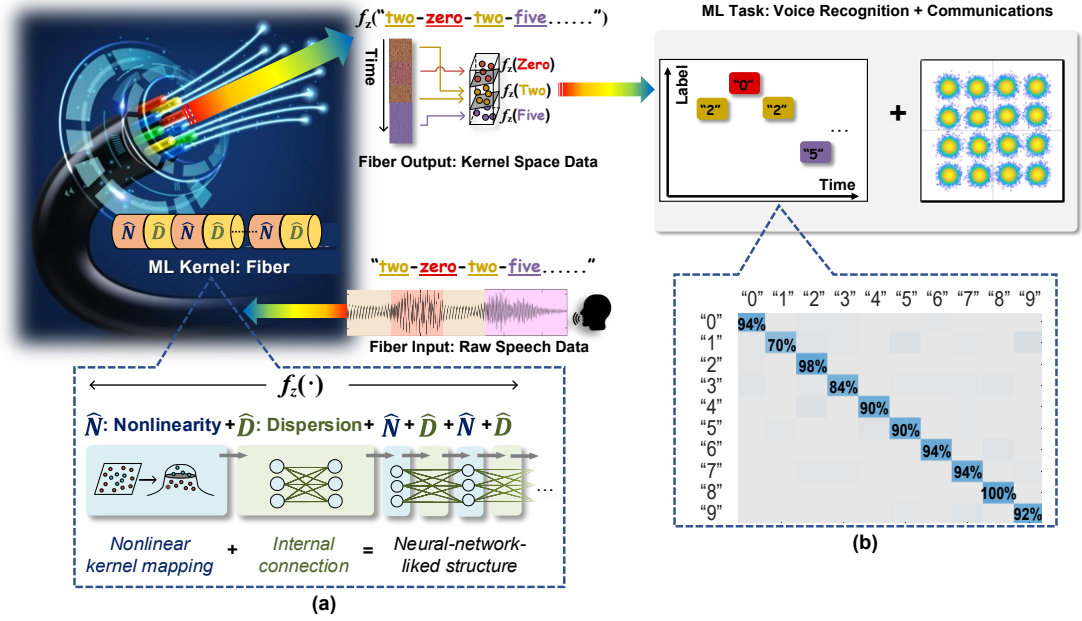

Fig. 1. The operation flow of the nonlinear fiber kernel function for ML voice recognition task, which can be integrated into optical fiber communication systems. (a) Demonstration of using fiber as a nonlinear kernel function, the CD and Kerr effects construct a neural-network-liked structure in fiber suitable for ML processing. (b) Confusion matrix of the ML voice recognition task.

*Fiber-optic System for Integrated Computation and Communication*

The structure of the fiber-optic transmission system can be divided into transmitter, channel, and receiver. The transmitter generates the signal source and modulates the information onto a light wave. The modulated optical signal is then guided through the fiber, which serves as the medium for light propagation, providing low-loss transmission and nonlinear kernel mapping. To mitigate signal degradation caused by attenuation and noise, amplifiers and filters are designed and placed within the fiber span to boost the signal strength.

After reaching the receiver, the optical signal is distorted during transmission. When the fiber is viewed as a low-loss channel for long-haul transmission, this distortion necessitates equalization to recover the original signal. As a result, the received signal is processed by digital signal processing (DSP) algorithms [10], where advanced techniques are employed to correct for CD, the Kerr effect, and other distortions.

*When the fiber is treated as a nonlinear kernel function for ML tasks, the received signal represents high-dimensional data that is inherently suitable for ML applications. Therefore, the received signal can be processed by simple ML algorithms such as ridge regression and applied to complex tasks like voice*

*recognition.*

Note that the communication and computation objectives can be achieved simultaneously by routing copies of the received signals to both processing flows, thereby integrating computation and communication. This ICAC system enables the long-distance transmission of data while simultaneously extracting and identifying features within the data, facilitating advanced applications such as optical semantic communication.

**EVALUATION SYSTEM AND RESULTS**

We numerically evaluate the system performance by constructing a 7×80 km fiber transmission link. The fiber used is standard single-mode fiber and is modeled using the SSFM with a step size of 1 km. The system configuration follows that described in Sec. II B. The original information is amplitude-modulated onto a 1550 nm optical carrier with a symbol rate of 32 Gbaud. The fiber launch power is set to 3 dBm, with an optical amplifier noise figure of 5 dB. The amplifier gain is optimized to counteract the attenuation.

To demonstrate the nonlinear mapping capability of the fiber kernel, we use a speech recognition task as the benchmark. The goal of the task is to classify spoken digits into their correct categories. The free-spoken-digit-dataset (FSDD) is chosen for this task. The spoken digits are feature-extracted and encoded onto the light wave. The ML algorithm used for classification is ridge regression, with a regularization parameter of 0.01.

The confusion matrix of FSDD recognition is presented in Fig. 1(b) and the overall accuracy for recognition is 90.6%. In comparison, a pure linear classifier [11] achieves an accuracy of 70.7%, indicating that the nonlinearity provided by the fiber kernel is crucial for the ML task. Additionally, comparisons with other schemes are made. A deep photonic reservoir computing system [11] achieves an accuracy of 92.4%, demonstrating that the nonlinear mapping capability of the proposed fiber kernel is comparable to advanced photonic computing strategies. Other hardware-based methods, such as the memristor cochlea scheme [12], achieve an accuracy of 92.0%. Software-based algorithms, like the echo state network [13], reach an accuracy of 87.5%, while advanced algorithm, such as recurrent neural networks [14], achieves an accuracy of 96.5%. The performance comparisons are summarized in TABLE I.

While the computation system successfully recognizes spoken digits, the communication system simultaneously transmits the signal with a bit error rate of $5.1 \times 10^{-5}$, aided by DSP. This demonstrates the effectiveness of the ICAC system.

TABLE I

PERFORMANCE COMPARISONS BETWEEN DIFFERENT COMPUTATIONAL SCHEMES

| Reference | Schemes | Accuracy | Reference | Schemes | Accuracy |
| --- | --- | --- | --- | --- | --- |
| This paper | Propose fiber kernel function | 90.6% | [12] | Cochlea based on memristor | 92.0% |
| [11] | Linear Classifier | 70.7% | [13] | Echo state network | 87.5% |
| [11] | Deep photonic reservoir computing | 92.4% | [14] | Recurrent neural network | 96.5% |

## CONCLUSIONS

In this paper, we present the integration of computation and communication processes within a fiber-optic system. A fiber kernel function for ML computing is developed by leveraging the inherent interactions between CD and Kerr effect of fiber-optic systems. Additionally, the optical fiber serves as a low-loss channel, ideal for long-haul communication. We numerically evaluate the performance of the fiber kernel function by conducting a simultaneous speech transmission and voice recognition task over a 7×80 km fiber transmission link. The recognition accuracy achieved is comparable to advanced photonic computing schemes, such as reservoir computing, demonstrating the nonlinear mapping capability of the fiber kernel function within the ICAC system.

## REFERENCES

[1] W. Zhang, D. Yang, C. Zhang, Q. Ye, H. Zhang, and X. Shen, "(Com)2Net: A Novel Communication and Computation Integrated Network Architecture," *IEEE Netw.*, vol. 38, no. 2, pp. 35–44, Mar. 2024, doi: 10.1109/MNET.2024.3355922.

[2] D. Wen, Y. Zhou, X. Li, Y. Shi, K. Huang, and K. B. Letaief, "A Survey on Integrated Sensing, Communication, and Computation," *IEEE Commun. Surv. Tutor.*, pp. 1–1, 2024, doi: 10.1109/COMST.2024.3521498.

[3] Y. He, G. Yu, Y. Cai, and H. Luo, "Integrated Sensing, Computation, and Communication: System Framework and Performance Optimization," *IEEE Trans. Wirel. Commun.*, vol. 23, no. 2, pp. 1114–1128, Feb. 2024, doi: 10.1109/TWC.2023.3285869.

[4] T. Zhou, F. Scalzo, and B. Jalali, "Nonlinear Schrödinger Kernel for Hardware Acceleration of Machine Learning," *J. Light. Technol.*, vol. 40, no. 5, pp. 1308–1319, Mar. 2022, doi: 10.1109/JLT.2022.3146131.

[5] T. Fu et al., "Optical neural networks: progress and challenges," *Light Sci. Appl.*, vol. 13, no. 1, p. 263, Sep. 2024, doi: 10.1038/s41377-024-01590-3.

[6] Q. Zhang and M. I. Hayee, "Symmetrized Split-Step Fourier Scheme to Control Global Simulation Accuracy in Fiber-Optic Communication Systems," *J. Light. Technol.*, vol. 26, no. 2, pp. 302–316, 2008, doi: 10.1109/JLT.2007.909861.

[7] J. Zhang et al., "Physics-Regulated Digital Backpropagation for Optical Fiber Systems with Imprecise Parameters," *IEEE Trans. Commun.*, pp. 1–1, 2024, doi: 10.1109/TCOMM.2024.3511694.

[8] A. Patle and D. S. Chouhan, "SVM kernel functions for classification," in *2013 International Conference on Advances in Technology and Engineering (ICATE)*, Jan. 2013, pp. 1–9. doi: 10.1109/ICAdTE.2013.6524743.

[9] U. Teğin, M. Yıldırım, İ. Oğuz, C. Moser, and D. Psaltis, "Scalable optical learning operator," *Nat. Comput. Sci.*, vol. 1, no. 8, pp. 542–549, Aug. 2021, doi: 10.1038/s43588-021-00112-0.

[10] J. Zhang, X. Yu, V. Bobrovs, X. Pang, O. Ozolins, and L. Zhang, "Analysis and Compensation of Nonlinear Dynamics in Optical Fiber Transmission with the Optoelectronic Reservoir Computing," in *2023 Photonics Global Conference (PGC)*, Aug. 2023, pp. 12–16. doi: 10.1109/PGC60057.2023.10344084.

[11] J. Zhang, L. Zhang, X. Pang, O. Ozolins, and X. Yu, "Accelerated Information Processing Based on Deep Photonic Time-Delay Reservoir Computing," *J. Light. Technol.*, vol. 42, no. 24, pp. 8739–8747, Dec. 2024, doi: 10.1109/JLT.2024.3438939.

[12] L. Cheng et al., "A bioinspired configurable cochlea based on memristors," *Front. Neurosci.*, vol. 16, Oct. 2022, doi: 10.3389/fnins.2022.982850.

[13] Y. Tanaka and H. Tamukoh, "Self-Organizing Multiple Readouts for Reservoir Computing," *IEEE Access*, vol. 11, pp. 138839–138849, 2023, doi: 10.1109/ACCESS.2023.3340311.

[14] R. Gayathri and K. Sheela Sobana Rani, "Implementation of Hybrid Deep Reinforcement Learning Technique for Speech Signal Classification," *Comput. Syst. Sci. Eng.*, vol. 46, no. 1, pp. 43–56, 2023, doi: 10.32604/csse.2023.032491.